\documentclass[sigconf]{acmart}

\usepackage{booktabs} 

\usepackage{balance}

\begin{document}
\title{The AIQ Meta-Testbed: Pragmatically Bridging Academic AI Testing and Industrial Q Needs}

\author{Markus Borg}
\orcid{XXX}
\affiliation{%
  \institution{RISE Research Institutes of Sweden AB\\Dept. of Computer Science, Lund University}
  \city{Lund}
  \country{Sweden}
}
\email{markus.borg@ri.se}

\renewcommand{\shortauthors}{M. Borg}

\begin{abstract}        
AI solutions seem to appear in any and all application domains. As AI becomes more pervasive, the importance of quality assurance increases. Unfortunately, there is no consensus on what artificial intelligence means and interpretations range from simple statistical analysis to sentient humanoid robots. On top of that, quality is a notoriously hard concept to pinpoint. What does this mean for AI quality? In this paper, we share our working definition and a pragmatic approach to address the corresponding quality assurance with a focus on testing. Finally, we present our ongoing work on establishing the AIQ Meta-Testbed.
\end{abstract}


%
%
%


\maketitle

\begin{abstract}        
AI solutions seem to appear in any and all application domains. As AI becomes more pervasive, the importance of quality assurance increases. Unfortunately, there is no consensus on what artificial intelligence means and interpretations range from simple statistical analysis to sentient humanoid robots. On top of that, quality is a notoriously hard concept to pinpoint. What does this mean for AI quality? In this paper, we share our working definition and a pragmatic approach to address the corresponding quality assurance with a focus on testing. Finally, we present our ongoing work on establishing the AIQ Meta-Testbed.

\end{abstract}
\section{Introduction} \label{sec:intro}

The number of AI applications is constantly growing. Across diverse domains, enterprises want to harness AI technology to explore the lucrative promises expressed by AI advocates. As AI becomes pervasive, there is inevitably a need to build trust in this type of software. Furthermore, critical AI is on the rise, i.e., applications will not be restricted to entertainment and games. AI is already fundamental in many business-critical applications such as ad optimization and recommendation systems. As the technology further evolves, many believe that safety-critical AI will soon become commonplace in the automotive~\cite{lipson2016driverless} and medical domains~\cite{Jiang230}. Other examples of critical AI, with other types of quality requirements, will be found in the finance industry and the public sector. Unfortunately, how to best approach Quality Assurance (QA) for AI applications remains an open question. 

A fundamental issue originates already in the terminology, i.e., the concept of ``AI quality''. First, there are several different definitions of AI, and their interpretations range from simple statistical analysis to the sentient humanoid robotics of the science fiction literature. Furthermore, AI appears to be a moving target, as what was considered AI when the term was coined in the 1950s would hardly qualify as AI today. Second, in the same vein, quality is a notoriously difficult aspect to pinpoint~\cite{walkinshaw_software_2017}. Quality is a multi-dimensional patchwork of different product aspects that influences the user's experience. Moreover, quality is highly subjective and largely lies in the eye of the beholder. Taken together, AI quality is a truly challenging concept to approach, i.e., a subjective mishmash of user experience regarding a type of technology with unclear boundaries that also change over time. There is a need for pragmatic interpretations to help advance research and practice related to AI quality -- we provide ours in Section~\ref{sec:defs}.

Contemporary AI solutions are dominated by Machine Learning (ML) and in particular supervised learning. A pragmatic first step would be to initially focus QA accordingly. As development of systems that rely on supervised learning introduces new challenges, QA must inevitably adapt. No longer is all logic expressed by programmers in source code instructions, instead ML models are trained on large sets of annotated data. Andrej Karpathy, AI Director at Tesla, refers to this paradigm of solution development as ``Software 2.0'' and claims that for many applications that require a mapping from input to output, it is easier to collect and annotate appropriate data than to explicitly write the mapping function.\footnote{bit.ly/3dKeUEH} As we embark on the AI quality journey, we argue that methods for QA of ``Software 2.0'' should evolve first -- we refer to this as \textit{MLware}.

The rest of this paper is organized as follows. Section~\ref{sec:bg} motivates the importance of MLware QA, elaborates on the intrinsic challenges, and presents closely related work. Section~\ref{sec:defs} introduces the working definitions used in our work on establishing the AIQ Meta-Testbed, which is further described in Section~\ref{sec:metatest}. Finally,  Section~\ref{sec:sum} concludes our position paper.

\section{Background and Related Work} \label{sec:bg}
Fueled by Internet-scale data and enabled by massive compute, ML using Deep Neural Networks (DNN), i.e., neural networks with several layers, has revolutionized several application areas. Success stories include computer vision, speech recognition, and machine translation. We will focus the discussion on DNNs, but many of the involved QA issues apply also to other families of ML, e.g., support vector machines, logistic regression, and random forests -- software that is not only coded, but also trained.

From a QA perspective, developing systems based on DNNs constitutes a paradigm shift compared to conventional systems~\cite{borg_safely_2019}. No longer do human engineers explicitly express all logic in source code, instead DNNs are trained using enormous amounts of historical data. A state-of-the-art DNN might be composed of hundreds of millions of parameter weights that is neither applicable for code review nor code coverage testing~\cite{salay_analysis_2018} -- best practices in industry and also mandated by contemporary safety standards. As long as ML applications are restricted to non-critical entertainment applications (e.g., video games and smartphone camera effects) this might not be an issue. However, when ML applications are integrated into critical systems, they must be trustworthy. 

The automotive domain is currently spearheading work on dependable ML, reflected by work on the emerging safety standard ISO/PAS~21448. DNNs are key enablers for vehicle environmental perception, which is a prerequisite for autonomous features such as lane departure detection, path planning, and vehicle tracking. While DNNs have been reported to outperform human classification accuracy for specific tasks, they will occasionally misclassify new input. Recent work shows that DNNs trained for perception can drastically change their output if only a few pixels change~\cite{azulay_why_2019}. The last decade resulted in many beaten ML benchmarks, but as illustrated by this example, there is a pressing need to close the gap between ML application development and its corresponding QA.

There are established approaches to QA for conventional software, i.e., software expressed in source code. Best practices have been captured in numerous textbooks over the years, e.g., by Schulmeyer~\cite{schulmeyer_handbook_1987}, Galin~\cite{galin_software_2003}, Mistrik \textit{et al.}~\cite{mistrik_software_2016}, and Walkinshaw~\cite{walkinshaw_software_2017}. 
Developers write source code that can be inspected by others as part of QA. As a complement, static code analysis tools can be used to support source code quality. Unfortunately, the logic encapsulated in a trained ML model cannot be targeted by QA approaches that work on the source code level. ML models in general, and DNN models in particular, are treated as black boxes. While there is growing interest in research on explainable AI~\cite{adadi_peeking_2018}, interpreting the inner workings of ML is still an open problem. This is a substantial issue when explainability is fundamental, e.g., when safety certification is required~\cite{greenyer_explainability_2019} or when demonstrating legal compliance~\cite{vogelsang_requirements_2019} (such as GDPR or absence of illegal discrimination in the trained model).

On the other hand, source code inspection and analysis are also not sufficient tools to perform QA of conventional software systems. During development, software solutions rapidly grow into highly complex systems whose QA rarely can be restricted to analysis -- although substantial research effort has been dedicated to formal methods~\cite{weyns_survey_2012} including formal verification in model-driven engineering~\cite{gonzalez_formal_2014}. In practice, software QA revolves around well-defined processes~\cite{herbsleb_software_1997,ashrafi_impact_2003} and a backbone of software testing. Software testing, i.e., learning about the system by executing it, is the quintessential approach to software QA~\cite{gelperin_growth_1988,orso_software_2014,kassab_software_2017}.

In the software engineering community, there is momentum on evolving practices to replace ad-hoc development of AI-enabled systems by systematic engineering approaches. A textbook by Hulten on ``Building Intelligent Systems''~\cite{hulten_building_2018} is recommended reading in related courses by K\"astner at Carnegie Mellon University~\cite{kastner_teaching_2020}
and Jamshidi at University of South Carolina. K\"astner also provides an annotated bibliography of related academic research\footnote{https://github.com/ckaestne/seaibib}, as does the SE4ML group at Leiden Institute of Advanced Computer Science\footnote{https://github.com/SE-ML/awesome-seml}, recently summarized in an academic paper~\cite{serban2020adoption}. Bosch \textit{et al.} recently presented a research agenda for engineering of AI systems~\cite{bosch_engineering_2020}, sharing what they consider the most important activities to reach production-quality AI systems. 

In recent years, numerous papers proposed novel testing techniques tailored for ML. Zhang \textit{et al.} conducted a comprehensive survey of 144 papers on ML testing~\cite{zhang_machine_2020}, defined as ``any activities designed to reveal ML bugs'' where an ML bug is ``any imperfection in a machine learning item that causes a discordance between the existing and the required conditions.'' Riccio \textit{et al.} conducted another secondary study, analyzing 70 primary studies on functional testing of ML-based systems~\cite{riccio_testing_2020}. The authors do not use the term ``bug'' for misclassifications, as any ML component will sometimes fail to generalize. We agree with this view, and avoid terms such as ML bugs, model bugs and the like when referring to functional inefficiencies of MLware. 


\section{AI Quality Assurance -- Working Definitions} \label{sec:defs}
As discussed in Section~\ref{sec:intro}, AI quality is a challenging concept to define. Consequently, QA for AI is at least as hard to specify. Still, we need a working definition to initiate efforts in this direction. In this section, we present the rationale behind our working definition of AI quality and AI quality assurance. Moreover, we introduce several related terms we use in collaborations with industry partners.

The original definition of AI from the 1950s is \textit{``the science and engineering of making intelligent machines''}. Unfortunately, this definition turns AI into a moving target, as expectations on what constitutes an intelligent machine change over time -- a computer program for logistics optimization in a warehouse would have been considered intelligent in the 1950s whereas it now could be part of an undergraduate computer science course. Since the term AI was introduced, it has often been used to refer to software solutions of the future, displaying increasingly human-like capabilities. The notation of ``intelligence'' is still common when referring to the gist of AI/ML applications, as in Hulten's textbook~\cite{hulten_building_2018}, but ideally we want a definition that remains the same over time.

We argue that the most useful view on AI is to consider it as the next wave of automation in the digital society. Extrapolating from the sequence 1) digitization, 2) digitalization, and 3) digital transformation~\cite{schallmo_history_2018}, we consider AI as the next enabling wave in the same direction -- allowing automation of more complex tasks than before. Our working definition of AI is \textit{``software that enables automation of tasks that normally would require human intelligence''}. While still imprecise, the definition is good enough for us to later define a delimited subset of AI that deserves our research focus.

Consulting the well-known textbook on AI by Russell and Norvig is one approach to explore the scope of AI~\cite{russell_artificial_2009}. The table of contents lists concepts such as searching, game playing, logic, planning, probabilistic reasoning, natural language processing, perception, robotics, and, of course, learning -- all important components when mimicking human intelligence. The textbook clearly shows that AI is more than ML. On the other hand, we argue that conventional software QA and testing can be applied to all AI techniques that are implemented in source code. Supervised and unsupervised learning, however, involves a transfer of control from source code to data. Research efforts on QA tailored for this new paradigm are what now would provide the highest return-on-investment. We need to focus on ML-enabled software -- we refer to this as \textit{MLware} for short. 

\begin{figure}
\begin{center}
\resizebox{.5\textwidth}{!}{\includegraphics{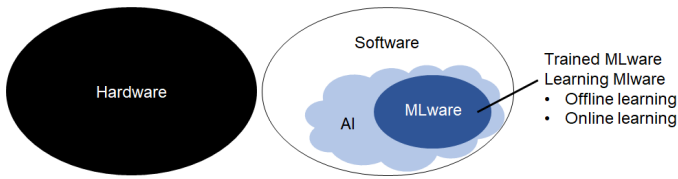}}
\caption{MLware in context.} \label{fig:mlware}
\end{center}
\end{figure}

Figure~\ref{fig:mlware} illustrates our view on MLware. The future of systems engineering will combine hardware and software components, but the software part needs to be differentiated. A subset of software represents the fuzzy area of AI. We accept that this subset is neither clear-cut nor consistent over time. MLware is a subset of AI that rely on supervised and/or unsupervised learning. All MLware is not made the same. From a QA perspective, we need to distinguish between \textit{trained MLware} that does not learn post deployment and \textit{learning MLware} that keeps improving as new experience is collected post deployment. Learning MLware can be further divided into offline learning (triggered re-training in batches) and online learning (continuous update of trained models).

One might wonder where Reinforcement Learning (RL) fits in our working definition of MLware. Currently, we exclude RL from MLware. The rationale is that in RL, the exploration and exploitation of the learning agent is implemented in source code. RL shares characteristics of both searching and automatic control. We posit that software testing approaches proposed for self-adaptive systems could be generalized to RL~\cite{cai_optimal_2002,mahdavi-hezavehi_systematic_2017}, and thus the best use of research resources is to focus on supervised and unsupervised learning -- the dominating types of ML in practical applications.

A well-cited experience report by Sculley and his Google colleagues presents the vast and complex infrastructure required for successful MLware~\cite{sculley_hidden_2015}. The authors describe this in terms of hidden technical debt of ML (cf. the lower part of Figure~\ref{fig:issues}). Building on this discussion, and the expression that ``data is the new oil'', our view is that data indeed fuels ML, but conventional source code is still in the driving seat, i.e., MLware is fueled by data and driven by code (cf. the upper part of Figure~\ref{fig:issues}). From this standpoint, it is obvious that conventional approaches to software QA remain essential in the new data-intensive paradigm of MLware. Moreover, just as software QA is dominated by software testing, we expect MLware QA to be dominated by MLware testing.

The phenomenon of software quality has been addressed in plentiful publications. Among other things, this has resulted in standardized software quality models such as ISO/IEC~25010. As MLware still is software, and certainly driven by source code, the existing quality models remain foundational. The sister standard, ISO/IEC~25012 Data Quality Model, adds a complementary data dimension to the quality discussion. As MLware is fueled by data, this standard is also highly relevant. Our working definition of AI quality is largely an amalgamation of the definitions provided by these two standards in the ISO/IEC 25000 series. 

As mentioned in Section~\ref{sec:bg}, there is no consensus in how to refer to issues resulting in MLware misclassifications. Bug is not a suitable term to cover all functional insufficiencies, given its strong connotation to source code defects. Still, we need a new similarly succinct term in the context of MLware. We propose \textit{snag} to refer to the difference between existing and required behaviors of MLware interwoven of data and source code. The root cause of a snag can be a bug either in the learning code or the infrastructure~\cite{zhang_machine_2020}, but it is often related to inadequate training data -- we call the latter phenomenon a \textit{dug}.

Figure~\ref{fig:issues} presents an overview of our perspective on issues detected in MLware. In the upper left, MLware is illustrated as a type of software that interweaves data (the fuel) and source code (at the helm) to produce output. If a discordance is observed, we call for a snag in the MLware fabric. Assuming that the requirements are valid and the observer interprets them correctly, root causes of snags include bugs and dugs as well as environment issues. The lower part of the figure illustrates the technical debt in machine learning as described by Sculley \textit{et al.}~\cite{sculley_hidden_2015}. Bugs can reside in the ML code (the white box), e.g., calling deprecated API methods or incorrect use of tensor shapes~\cite{humbatova2019taxonomy}. On the other hand, there might also be bugs in the rest of the infrastructure. While the illustrated technical debt revolves around data, all gray boxes will also depend on source code, from small exploratory scripts to mature open source libraries -- and the large systems enabling MLware operations~\cite{hulten_building_2018}.

\begin{figure*}
\begin{center}
\includegraphics{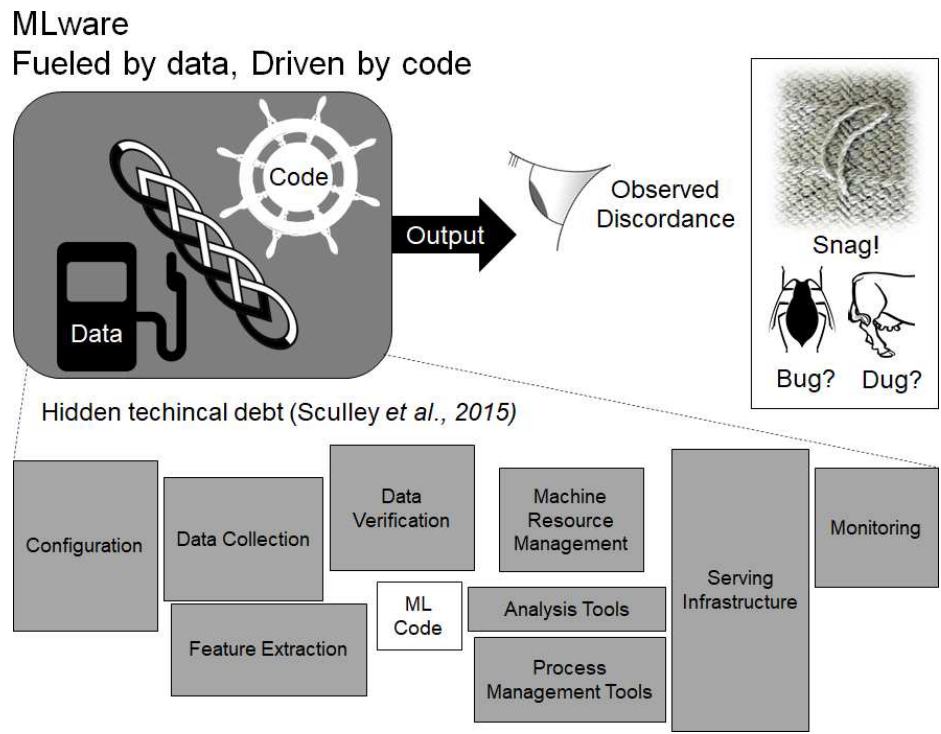}
\caption{MLware interwoven by data and code. Observed discordances in the output (snags) can originate in source code defects (bugs) or data inadequacies (dugs).} \label{fig:issues}
\end{center}
\end{figure*}

To summarize this section, our position is that research on QA for AI would benefit from adhering to the definitions presented in Table~\ref{tab:defs}.

\begin{table*}[]
\caption{Working definitions of key terms related to the AIQ Meta-Testbed.}
\begin{tabular}{|p{1.5cm}|p{5cm}|p{6cm}|}
\hline
\multicolumn{1}{|c|}{\textbf{Term}}                                     & \multicolumn{1}{c|}{\textbf{Definition}}                                                                                                                                                         & \multicolumn{1}{c|}{\textbf{Comments}}                                                                                                                                                                                                                                                                                                                                                         \\ \hline
AI                                                         & A subset of software that automates tasks that normally would require human intelligence.                                                                             & {MLware, interwoven by data and source code, is the most precise term to describe our research interest. On the other hand, AI is a dominant term in industry and news media. We propose a pragmatic sacrifice of scientific preciseness in favour of industrial and societal relevance. 
In practice, we treat AI as synonymous with MLware in discussions with clients.} \\ \cline{1-2}
MLware                                                             & A subset of AI that, fueled by data, realizes functionality through supervised and/or unsupervised learning.                                       &                                                                                                                                                                                                                                                                                                                                                                                       \\ \hline
\begin{tabular}[c]{@{}l@{}}MLware\\ Testing\end{tabular}       & Any activity that aims to learn about MLware by executing it.                                                                                                                           & The typical goal of testing is detecting differences between existing and required behavior~\cite{ammann2016introduction}. Other possible testing goals include exploratory testing and compliance testing.                                                                                                                                                        \\ \hline
AI Quality                                                     & The capability of MLware to satisfy stated and implied needs under specified conditions while the underlying data satisfy the requirements specific to the application and its context. & MLware combines data and conventional source code, thus we propose the amalgamation of corresponding quality definitions from the IEC/ISO 25000 series. Our proposal is in line with discussions by Felderer \textit{et al.} in the context of testing data-intensive systems~\cite{Felderer2019}.                                                                                                                                                                                                                                       \\ \hline
\begin{tabular}[c]{@{}l@{}}AI Quality\\ Assurance\end{tabular} & Any systematic process to provide confidence that the desired AI Quality is maintained.                                & QA encompasses many activities throughout the product lifecycle. However, in current AI discussions with clients, we primarily interpret it as MLware testing.                                                                                                                                                                                                                         \\ \hline
Snag                                                           & Any imperfection in MLware that causes a discordance between the existing and the required conditions.                                             & There is an ongoing discussion in the research community about how to refer to MLware misclassifications~\cite{riccio_testing_2020}. We argue against using the term bug whenever there is unexpected output. Instead, we propose calling it a snag in the MLware fabric.                                                                                     \\ \hline
Bug                                                            & A source code defect that causes a discordance between the existing and the required conditions.                                                                                                   & The term bug has a firmly established meaning, thus we suggest restricting its use to source code. As MLware is driven by code, bugs can cause snags.                                                                                                                                                                                                                               \\ \hline
Dug                                                            & A data inadequacy that causes a discordance between the existing and the required conditions.                                                                                                      & With bugs reserved for source code defects, we need a novel expression for the data counterpart. The new term must be a worthy match for the succinct ``bug''. Currently, we call them ``dugs''.                                                                                                                                                                                               \\ \hline
\end{tabular}
\label{tab:defs}
\end{table*}

\section{AIQ -- An AI Meta-Testbed} \label{sec:metatest}
Based on the working definitions in Section~\ref{sec:defs}, we plan to support AI QA by establishing an AI meta-testbed. A testbed is a venue that provides a controlled environment to evaluate technical concepts. Under current circumstances, in the middle of the ongoing AI boom\footnote{Well aware of the two previous ``AI winters'', periods with less interest and funding due to inflated expectations.}, we believe that the establishment of a testbed for testing MLware testing would be the most valuable contribution to AI QA. Assessing the effectiveness of different testing techniques in a controlled setting is not a new idea~\cite{basili_comparing_1987}, neither is the concept of testing test cases~\cite{zhu_systematic_2018} -- but a testbed dedicated to MLware testing is novel. We call it the AIQ Meta-Testbed\footnote{metatest.ai}.

Successful MLware development requires a close connection to the operational environment. The same need has shaped software development at Internet companies, resulting in DevOps -- a combination of philosophies, practices, and tools to reduce the time between development and operations while preserving quality~\cite{erich2017qualitative}. Key enablers are Continuous Integration and Deployment (CI/CD). DevOps that emphasize MLware development is often referred to as MLOps~\cite{karamitsos2020applying}, effectively adding Continuous Training (CT) to the mix. The focus on continuousness is stressed in illustrations by the infinity symbol.

Trust is fundamental for a successful product or service embedding MLware. In 2019, an expert group set up by the European Commission published ethics guidelines for trustworthy AI\footnote{ec.europa.eu/digital-single-market/en/news/ethics-guidelines-trustworthy-ai}. As part of the guidelines, seven key requirements are introduced. Table~\ref{tab:EU} shows a mapping between the EU requirements and the testing properties identified in the survey by Zhang \textit{et .al.}~\cite{zhang_machine_2020}. Our preliminary analysis indicates that all but one requirement has (to some extent) been targeted by academic research. Thus, we believe the time is right for systematic meta-testing in an MLOps context. 

\begin{table*}[]
\caption{Mapping the EU requirements for trustworthy AI and the testing properties targeted by publications on MLware testing as identified by Zhang \textit{et al.}~\cite{zhang_machine_2020}. Gray cells show functional testing, i.e., the scope of Riccio \textit{et al.}'s secondary study~\cite{riccio_testing_2020}}
    \begin{center}
        \includegraphics{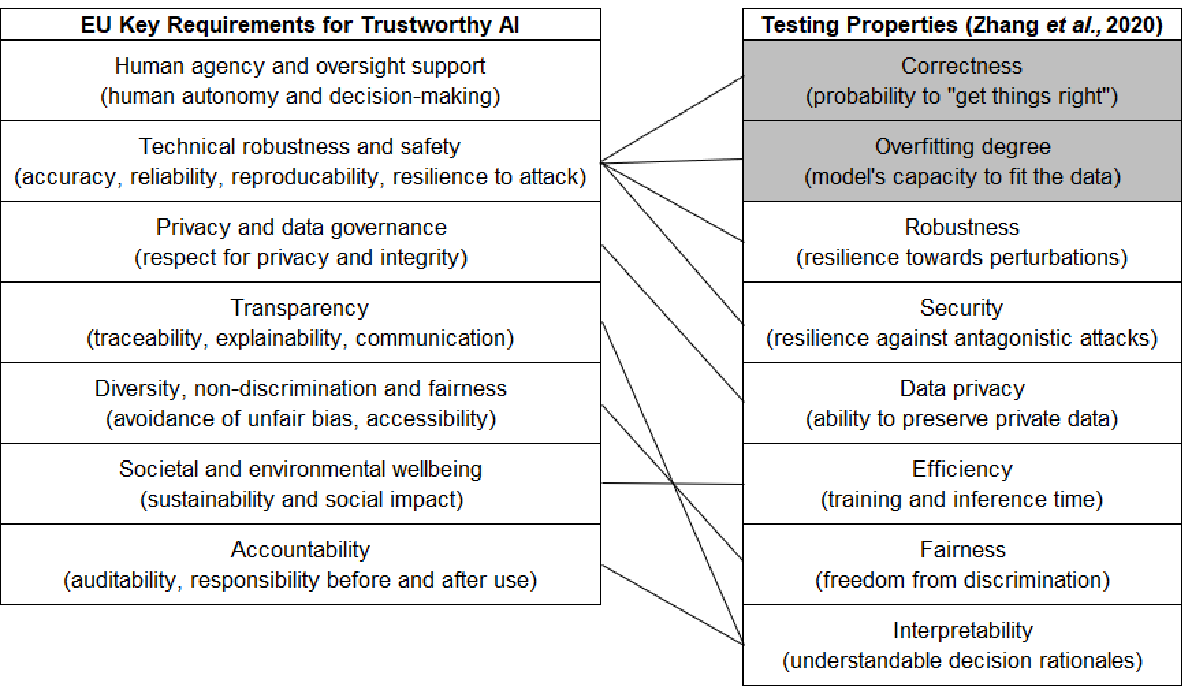}
 \label{fig:issues}
    \end{center}
    \label{tab:EU}
\end{table*}

\begin{figure*}
\begin{center}
\includegraphics{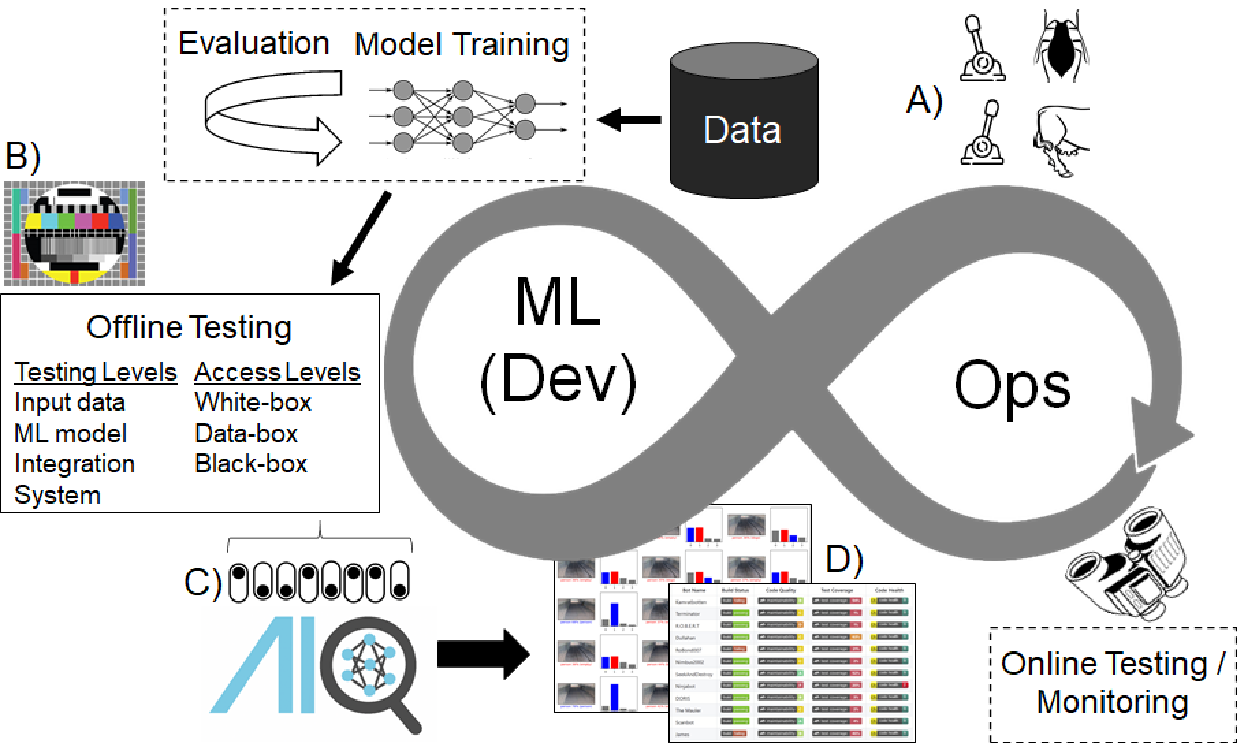}
\caption{The AIQ Meta-Testbed in the MLOps context. We will focus on providing A) fault-injection, B) test input generation for offline testing, C) a control panel for toggling offline testing techniques, and D) presenting the results in dashboards.}
\label{fig:aiq}
\end{center}
\end{figure*}

Figure~\ref{fig:aiq} presents an overview of the AIQ Meta-Testbed in the MLOps context. We will set up a contemporary MLOps pipeline to allow controlled experiments in the lab while still providing an environment relevant to industry practice. Test automation is the backbone of MLOps, and MLware testing occurs in several phases during the MLware engineering lifecycle~\cite{zhang_machine_2020} (cf. the textboxes in Figure~\ref{fig:aiq}). First, the standard practice during model training is to split data into training, validation, and test subsets. We refer to this type of ML model testing as evaluation. Second, offline MLware testing occurs prior to deployment -- conducted on different testing levels (input data, ML model, integration, system) and with varying access levels of the MLware under test (white-box, data-box, black-box) as defined by Riccio \textit{et al.}~\cite{riccio_testing_2020}. Third, online MLware testing occurs after deployment. Common examples include A/B testing and runtime monitoring to detect distributional shifts.

The AIQ Meta-Testbed will primarily focus on offline MLware testing (the solid-border textbox in Figure~\ref{fig:aiq}). We plan to enable meta-testing by providing a \textit{control panel} for toggling testing techniques (C) in Figure~\ref{fig:aiq}) corresponding to the testing properties in Table~\ref{tab:EU}, controlled \textit{fault-injection} (A) (e.g., bug/dug injection, hyperparameter changes, mutation operators) and state-of-the-art \textit{test input generation} (B) (e.g., search-based testing, GAN-based synthesis, metamorphic relations, and adequacy-driven generation). The results from both MLware testing and meta-testing will be presented in dashboards (D).

Extrapolating from the publication trends reported in the recent secondary studies~\cite{zhang_machine_2020,riccio_testing_2020}, there will be an avalanche of MLware testing papers in the next years. Staying on top of the research will become a considerable challenge and for practitioners with limited experience in reading academic papers, the challenge will be insurmountable -- motivating the need to create an overview and shortlisting the most promising techniques.

Activities at the AIQ Meta-Testbed will include external replications of studies on MLware testing. By performing controlled meta-testing of the shortlisted techniques,  we will be able to provide evidence-based recommendations on what techniques to use and in which contexts. The controlled environment of the AIQ Meta-Testbed will enable exploration of applied research questions, such as:
\begin{itemize}
    \item Which contextual factors influence the MLware test effectiveness the most? 
    \item Which proposed MLware testing techniques scale to very large DNNs?
    \item How to best integrate MLware testing in an MLOps pipeline?
    \item What should be done to limit test maintenance in an MLware testing context?
    \item After observing a snag, how to support the subsequent root cause analysis?
\end{itemize}

\section{Summary and Concluding Remarks} \label{sec:sum}
AI is becoming a pervasive subset of software, thus the elusive concepts of AI quality and QA are increasingly important. We argue that pragmatic interpretations are needed to advance the field, and introduce a working definition of MLware as a subset of software within AI that realizes functionality through machine learning by interweaving data and source code. Furthermore, we define AI quality as ``the capability of MLware to satisfy stated and implied needs under specified conditions while the underlying data satisfy the requirements specific to the application and its context''. We recommend that AI QA first and foremost should be interpreted as MLware testing and that the term bug shall be reserved for source code defects -- instead we propose ``snag'' to refer to observed discordances in the MLware fabric. Finally, we present the AIQ Meta-Testbed -- bridging academic research on MLware testing and industrial needs for quality by providing evidence-based recommendations based on replication studies in a controlled environment.

\section*{Acknowledgements}
This work was funded by Plattformen at Campus Helsingborg, Lund University.

\bibliographystyle{ACM-Reference-Format}
\bibliography{aiq}

\end{document}